\documentstyle[psfig]{mn}

\newif\ifAMStwofonts

\include{nfss.tex}

\newcommand{\Msolar}{\mbox{$M_{\odot}\,$}}
\newcommand{\halpha}{\mbox{${\rm H}\alpha$}}
\newcommand{\kms}{\mbox{${\rm \,km\,s}^{-1}$}}

\pubyear{1998}

\begin{document}

\title{The fraction of double degenerates among DA white dwarfs.}
\author[P.F.L. Maxted \& T.R. Marsh]
       {P.F.L. Maxted \& T.R. Marsh \\
 University of Southampton, Department of Physics \& Astronomy,
        Highfield, Southampton, S017 1BJ, UK}
\maketitle

\label{firstpage}

\begin{abstract}
 We present the results of a radial velocity survey designed to measure the
fraction of double degenerates among DA white dwarfs. The narrow core of the
H$\alpha$ line was observed twice or more for 46 white dwarfs yielding radial
velocities accurate to a few \kms.  This makes our survey the most
sensitive to the detection of double degenerates undertaken to date. We found
no new  double degenerates  in our sample, though \halpha\ emission from
distant companions is seen in two systems. Two stars known to be double
degenerates prior to our observations are included in the analysis.  We find a
95\% probability that the fraction of double degenerates among DA white dwarfs
lies in the range $[0.017,0.19]$. 
\end{abstract} 

\section{Introduction}
 White dwarfs are the ``fossil remnants'' of low and intermediate mass stars
so their study might well be named ``paleaoastrophysics'' - the study of the
history of our Galaxy through the fossil remnants of its population. Key
parameters for any population of stars are the binary fraction and the period
and mass ratio distributions of the binaries. These parameters
will be different among white dwarfs compared to their parent population of
main-sequence stars because the large size reached by a star during its
red-giant phase leads to interactions with nearby companions. The results of
such interactions are complicated but certainly lead to ejection of some or all
of the red giant envelope at the expense of orbital angular momentum. This
``common-envelope'' phase will occur for brown-dwarf and massive planetary
companions (Nelemans \& Tauris, 1998) e.g. 51\,Peg\,B (Mayor \& Queloz 1995),
as well as for stellar companions. If two common-envelope phases or other mass
transfer episodes occur (e.g. an Algol-like phase, Sarna et~al., 1996) the
result may well be a short-period white dwarf\,--\,white dwarf binary -- a
double-degenerate star (DD). In short period DDs the loss of orbital angular
momentum through gravitational radiation causes the binary to coalesce. The
physics of this merging is complicated but it is reasonable to assume that if
the total mass of the binary exceeds the Chandrasekhar limit, some sort of
violent explosion may occur. Thus, DDs have been proposed as the source of
Type~Ia supernovae (Iben et~al., 1997). 

There have been several surveys for double-degenerates (Saffer et~al. 1998;
Bragaglia 1990; Robinson \& Shafter 1987; Foss et~al., 1991). The observing
strategies used have tended to be optimized for the detection of very short
period systems ($P <$ few hours). These very short period systems will merge
within a Hubble time or less through the loss of angular momentum via
gravitational wave radiation. By measuring the space density of these very
short period systems, it is possible, in principle, to establish whether
or not there are enough such systems to account for the observed rate of Type
Ia supernovae (SNe~Ia) in our Galaxy. However, these very short period systems
may comprise as few as 1/570 of the white-dwarf population; (Saffer et~al.,
1998), so the number of systems surveyed would  need to be increased by at
least an order-of-magnitude before surveys of this type can be used to draw
any firm conclusions regarding the space-density of SNe~Ia progenitors. 

 An alternative approach is to study the longer period systems. Although they
do not provide a direct measure of the current rate of Galactic SNe~Ia
 they are predicted to be the most common type of DD. As such, they
provide a good means of  testing models of binary star evolution leading to
DDs. These models also predict the properties of other types of binaries
arising from common-envelope evolution e.g. cataclysmic variables, black hole
binaries, binary pulsars etc. Double-degenerates do not suffer from poorly
understood phenomena such as magnetic-wind braking and mass transfer which
make it so difficult to test models of binary evolution using observations of
those binary stars which show these effects. Also, the selection effects
associated with searches for DDs  are, in principle, straightforward to
calculate. Thus, DDs are the best objects to study if we are to critically
test models of binary star evolution.

 In this paper we report the results of a radial velocity survey of DA
white-dwarfs. We then use the results of this survey to show there is a 95\%
probability that the fraction of double degenerates among DA white dwarfs lies
in the range $[0.017,0.19]$.

\section{Observations and reductions} 
\setcounter{table}{1}
\begin{table}
\caption{\label{ResultsTable} Summary of the radial velocity measurements for
our sample of DA white dwarfs.}
\begin{tabular}{lrrrrrrr}
WD &  \multicolumn{1}{l}{N}&\multicolumn{1}{l}{Mean RV}&
\multicolumn{1}{l}{$\chi^2$}&  \multicolumn{1}{l}{$\log_{10}p$}&
\multicolumn{1}{l}{Mass} \\
   & &\multicolumn{1}{l}{(\kms)}& & & 
\multicolumn{1}{l}{(\Msolar)} \\
0047-524 & 5 & 35.8 $\pm$ 0.7 & 5.89 & -0.68&    $0.50^3$\\
0101+048 & 3 & 63.7 $\pm$ 1.5 & 0.33 & -0.07&    \\
0226-329 & 2 & 27.0 $\pm$ 2.4 & 4.17 & -1.38&    \\
0227+050$^{\star}$ & 4 & 18.8 $\pm$ 0.6 & 1.98 & -0.24&    $0.48^1$\\
0310-688 & 14 & 65.9 $\pm$ 0.1 & 11.15 & -0.22&  $0.63^3$\\
1149+057 & 3 & 2.0 $\pm$ 3.4 & 0.11 & -0.02&     \\
1210+140 & 2 & 69.7 $\pm$ 5.6 & 1.42 & -0.63&    \\
1233-164 & 3 & 66.2 $\pm$ 4.3 & 0.99 & -0.21&    \\
1310-305 & 4 & 40.1 $\pm$ 0.9 & 2.05 & -0.25&    \\
1314-153 & 5 & 108.8 $\pm$ 0.8 & 9.31 & -1.27&   \\
1327-083 & 5 & 44.7 $\pm$ 0.3 & 2.96 & -0.25& $0.54^1$   \\
1348-273 & 3 & 62.4 $\pm$ 2.7 & 0.36 & -0.08&    $0.50^3$\\
1425-811 & 5 & 34.6 $\pm$ 0.7 & 3.85 & -0.37&    $0.68^2$\\
1550+183 & 3 & 15.8 $\pm$ 1.8 & 0.32 & -0.07&    \\
1616-591 & 4 & 9.5 $\pm$ 1.5 & 1.45 & -0.16&     \\
1619+123 & 4 & 22.4 $\pm$ 1.5 & 3.15 & -0.43&    \\
1620-391 & 7 & 46.4 $\pm$ 0.2 & 5.10 & -0.28&    $0.66^3$\\
1659-531 & 3 & 51.3 $\pm$ 1.0 & 0.11 & -0.02&    \\
1716+020 & 5 & -15.7 $\pm$ 0.7 & 2.07 & -0.14&   $0.43^2$\\
1743-132 & 5 & -68.5 $\pm$ 1.1 & 0.71 & -0.02&   $0.47^2$\\
1826-045 & 4 & 1.9 $\pm$ 0.8 & 2.24 & -0.28&     \\
1827-106 & 3 & -31.1 $\pm$ 3.1 & 0.71 & -0.15&   \\
1840+042 & 3 & 5.8 $\pm$ 1.1 & 0.67 & -0.15&     \\
1840-111 & 4 & -5.9 $\pm$ 0.9 & 1.00 & -0.10&    \\
1845+019$^{\star}$ & 5 & -29.9 $\pm$ 0.7 & 7.10 & -0.88&   $0.51^1$\\
1845+019B & 5 & -49.2 $\pm$ 0.8 & 9.61 & -1.32&  \\
1914-598 & 4 & 74.9 $\pm$ 0.9 & 0.10 & 0.00&     \\
1919+145 & 5 & 53.0 $\pm$ 0.6 & 6.34 & -0.76&    \\
1943+163 & 4 & 36.5 $\pm$ 1.2 & 0.64 & -0.05&    $0.49^1$\\
2007-303 & 5 & 75.4 $\pm$ 0.2 & 16.46 & -2.61&   $0.51^1$\\
2014-575 & 4 & 41.1 $\pm$ 1.5 & 0.62 & -0.05&    $0.54^4$\\
2035-336 & 5 & 20.9 $\pm$ 0.8 & 1.80 & -0.11&    \\
2039-202 & 7 & 1.8 $\pm$ 0.3 & 5.60 & -0.33&     $0.56^1$\\
2039-682$^{\star}$ & 3 & 55.0 $\pm$ 2.6 & 2.54 & -0.55&    $0.87^1$\\
2058+181 & 3 & -43.0 $\pm$ 2.7 & 1.47 & -0.32&   \\
2105-820 & 5 & 42.6 $\pm$ 0.8 & 5.19 & -0.57&    \\
2115-560 & 5 & 9.8 $\pm$ 0.7 & 8.05 & -1.05&     $0.66^1$\\
2149+021 & 4 & 32.0 $\pm$ 0.5 & 1.83 & -0.22&    $0.61^4$\\
2151-015$^{\star}$ &  5 &   35.7 +/-  1.2 & 11.91 & -1.74& \\
2151-015B &  5 &    7.0 +/-  1.6 &  1.70 & -0.10& \\
2151-307 & 3 & 55.2 $\pm$ 3.6 & 0.95 & -0.21&    \\
2159-754 & 2 & 153.2 $\pm$ 3.3 & 0.00 & -0.01&   \\
2211-495$^{\star}$ & 5 & 37.5 $\pm$ 1.3 & 1.90 & -0.12&    \\
2251-634 & 5 & 32.5 $\pm$ 0.7 & 3.08 & -0.26&    \\
2326+049$^{\star}$ & 5 & 44.7 $\pm$ 0.7 & 7.91 & -1.02&    \\
2333-165 & 4 & 72.3 $\pm$ 0.4 & 0.18 & -0.01&    \\
2351-335 & 5 & 51.0 $\pm$ 0.5 & 7.57 & -0.96&    \\
2359-434$^{\star}$ & 8 & 43.6 $\pm$ 0.6 & 15.14 & -1.46&   $0.97^1$\\
\end{tabular}

$^{\star}$ See text for notes concerning this object. \newline
1.\, Bergeron et~al., 1992 2.\,Bergeron et~al., 1995
3.\,Bragaglia et~al., 1995 4.\,Finley et~al. 1997
\end{table}

Our observations were obtained with the 3.9m Anglo-Australian Telescope at
Siding Spring, Australia over two observing runs, 1997 August 15-17 and 1998
June 3-5. Conditions were generally good with a total of 1.5 nights lost to
cloud, typical seeing of 1--2\,arcsec  and a final night with sub-arcsecond
seeing throughout. The RGO spectrograph with the 82cm camera and
1200l\,mm$^{-1}$ grating was combined with a 1\,arcsec wide slit to obtain a
resolution of 0.7\AA\ at \halpha. The detector for the first observing run was
a TEK CCD with 1024 pixels covering 0.235\AA\ each.  The detector for the
second run was a MIT-LL CCD. We used only a portion of this detector and
on-chip binning over two pixels in the spectral direction to give 1536 pixels
of 0.293\AA\ each. Exposure times were typically 900s or 1200s giving a
signal-to-noise ratio of 30 or more in the continuum for a white dwarf of
around 14th magnitude. Small scale sensitivity variations were removed using
observations of an internal tungsten calibration lamp and images of the
twilight sky were used to remove the variation of slit throughput in the
spatial direction. Bias images over both detectors show no signs of any
structure so a constant bias level determined from the median value in the
over-scan region was subtracted from all the images.  

 Extraction of the spectra from the images was performed automatically using
optimal extraction to maximize the signal-to-noise of the resulting spectra
(Horne, 1986). Every spectrum was bracketed by observations of the internal
copper-argon arc lamp at the position of the star. The arcs associated with
each stellar spectrum were extracted using the profile determined for
the stellar image to avoid possible systematic errors due to tilted spectra.
The wavelength scale was determined from a fourth-order polynomial fit to
measured arc-line positions. The 1997 spectra were normalized using a linear
fit to the continuum region either side of broad \halpha\ line. Some of the
1998 spectra show strong vignetting over $\approx \frac{1}{3}$ of the blue end
of the spectra due to a shutter falling partially into the spectrograph beam.
Therefore, all the 1998 spectra have been normalized using a linear fit to the
continuum to the red of H$\alpha$ only. To determine the amount of vignetting
in the affected spectra we calculated the ratio of these spectra to either an
unvignetted spectrum of the same star or a multiple Gaussian fit to the
unvignetted portion of the spectrum. A smooth function was then used to model
this vignetting and this was used to correct the original spectrum. Note that
this procedure only affects the regions of the spectrum blue-wards of the core
and will not affect the measurement of radial velocities.

 Spectra were obtained for a total 49 objects taken from the catalogue of
McCook \& Sion (1998). All the objects observed are brighter than about 15th
magnitude and have a `DA' catalogue spectral type. Where a sub-class is given
with the spectral type, only objects with sub-types 2 or greater were
included.

\subsection{The observing strategy.}
\begin{figure}
\leavevmode\centering{
\psfig{file=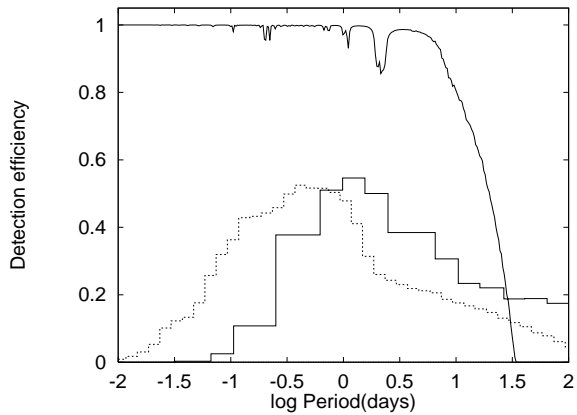,width=0.45\textwidth} }
\caption{\label{DetectionEfficiencyFig}Typical detection efficiency for our
observing strategy (solid line) and the theoretical period distributions for
DDs of Han (histogram, solid lines) and Iben et~al. (histogram, dotted lines).}
\end{figure}

We attempted to obtain 4 spectra of each white dwarf such that intervals
between spectra were approximately 20 minutes, 4 hours and 1\,--\,2 days, e.g.
\begin{itemize}
\item{a pair of consecutive spectra;}
\item{one spectrum, 3--6 hours later;}
\item{one spectrum the following night.} 
\end{itemize}
There are very few periods which fit this sampling in such a way as to avoid
detection, particularly given our high radial velocity accuracy. Therefore,
the detection efficiency is extremely high across a wide range of periods.
This is shown in Fig.~\ref{DetectionEfficiencyFig} where we plot the fraction
of binaries satisfying our detection criterion as a function of period for a
typical observing sequence and radial velocity accuracies. The effects of
randomly oriented orbits and random zero phases have been included and 
masses of 0.5\Msolar for both components have been assumed. We also plot the
normalized theoretical period distributions for DDs of Iben et~al. (1997) 
and Han (1998) for comparison. 

\subsection{Measurement of radial velocities.}
 We used least-squares fitting of four Gaussian profiles to measure radial
velocities from our spectra. We used an automatic procedure to yield objective
results as far as possible. For each star we first used a simultaneous fit to
 all the spectra of that star to find the optimum width and depth of the
Gaussian profiles. These parameters were then fixed and a second fit to each
of spectra was used to measure the radial velocities.  The radial velocity of
each spectrum and the coefficients of a second-order polynomial to model the
continuum were the only parameters optimized in this second fitting procedure.
This method gives satisfactory fit to almost all the spectra with no further
effort. The few exceptions to this rule and other objects of interest are
noted in the following section. All the measured radial velocities used in
this study are given in Table~\ref{RVTable}.

\subsection{Notes on individual systems.}  
\begin{description}
\item[\bf WD\,0107-342]{This is not included in our sample as our single
spectrum shows a strong HeI 6676 absorption line which suggests that this is 
an sdB star.}
\item[\bf WD\,1845+019]{The core of \halpha\ shows a weak, sharp emission line
slightly to the blue of the absorption core of \halpha\ (Fig.~\ref{WD1845Fig}).
We used an additional Gaussian profile with an independent radial velocity in
the fitting process to model this emission. Neither  component shows any
radial velocity variation and the mean offset between the components is around
20\kms, though this is somewhat uncertain as the emission and absorption cores
are not fully resolved. Nevertheless, we have used the mass (0.57\Msolar) and
surface gravity ($\log g = 7.84$) of this star measured by Finley et~al.
(1997)  to predict a gravitational redshift for this star of 24\kms, which is
in good agreement with our measured offset. We conclude that the emission
arises from a cool companion star in a long period orbit around this white
dwarf. The radial velocity of the emission component is listed in
Table~\ref{ResultsTable} as WD\,1845+019\,B. The radial velocity measurements
for WD\,1845+019 are included in our analysis, but those of its companion are
not.} 
\begin{figure} \leavevmode\centering{
\psfig{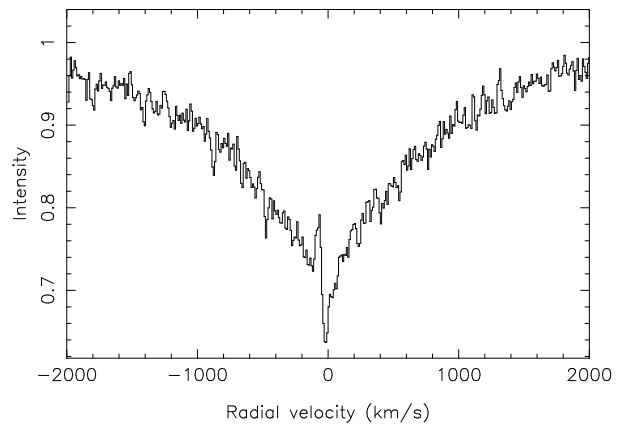} }
\caption{\label{WD1845Fig} The average of our 5 spectra of WD\,1845+019
showing the emission from the cool companion.} \end{figure} 
\item[\bf WD\,1953-011]{Our single spectrum of this object shows `shoulders'
$\sim 450\kms$  either side of the the broad core of the \halpha\ line
(Fig.~\ref{WD1953Fig}). Note that these are {\it not} the features noted by
Koester et~al, 1998. The latter occur at a much smaller scale ($\sim 100\kms$)
and are the cause of the odd shaped core of the \halpha\ line. A degree of
smoothing has been applied to the spectrum shown in Fig.~\ref{WD1953Fig} to
show more clearly that our core profile is consistent with the Zeeman triplet
seen in the spectra of Koester et~al. (their Fig.~7) which implies a mean
magnetic field of $93\pm5$\,kG. If the features at $\pm 450\kms$ are  also due
to Zeeman splitting, this implies a mean field strength of $\sim 500$\,kG.
This would require an unusual magnetic field geometry for a single white
dwarf, which might suggest that WD\,1953-011 is a binary composed of two
magnetic white dwarfs. Further observations, particularly circular
polarization measurements, are required to address these issues. } 
\begin{figure} \leavevmode\centering{
\psfig{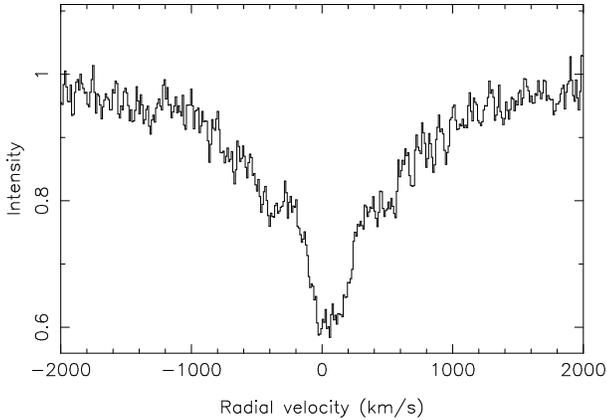} }
\caption{\label{WD1953Fig} Our single spectrum of WD\,1953-011.} \end{figure} 
\item[\bf WD\,2039-682]{The core of \halpha\ is noticeably broader and
shallower than those of white dwarfs of similar spectral type (Fig.~4). One of
the stars for which we have several high signal-to-noise spectra,
WD\,0310-688, has a very similar temperature and gravity to WD\,2039-682
(Bragaglia et~al., 1995). We applied a simple rotational broadening function
to our average spectrum of WD\,0310-688 assuming a linear limb darkening
coefficient of 0.15 (Koester et~al., 1998). By comparing the average spectrum
of WD\,2039-682 to the broadened spectra of WD\,0310-688 for various values of
the projected rotational velocity, $V\sin i$, we estimate $V\sin i= 82\pm
5\kms$, which agrees well with the estimate of Koester et~al. ($V\sin i= 78\pm
6\kms$). The resolution of our spectra is lower than theirs but our
signal-to-noise ratio is much higher. The very good fit obtained by this
method strongly suggests this is a genuine rapidly rotating white-dwarf. }
\begin{figure} \leavevmode\centering{
\psfig{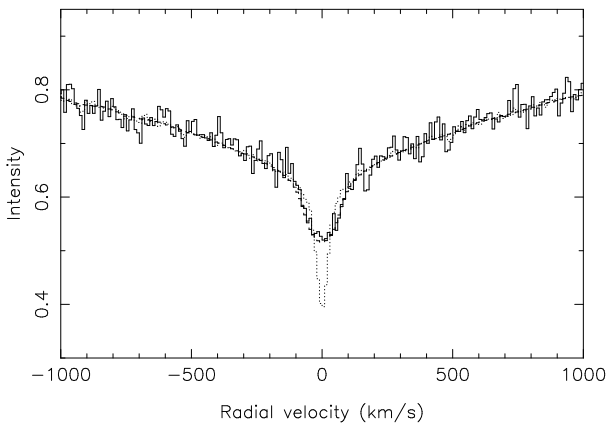} }
\caption{\label{WD2039Fig} The average spectrum of WD\,2039-682 (solid line),
the average spectrum WD\,0310-688 before applying any broadening (dotted line)
and after  applying rotational broadening for the measured value of $V\sin i$
(dashed line).} \end{figure} 
\item[\bf WD\,2151-015]{One of the spectra of this star shows a strong
emission line slightly to the blue of the absorption core which, with
hindsight, is also present at a much weaker level in at least two other
spectra (Fig.~\ref{WD2151Fig}). By fixing the width of the emission line from
a fit to the spectrum in which it is strongest, we were able to measure
independent velocities for the emission and absorption components. The mean
radial velocity of the emission component is given in Table~\ref{ResultsTable}
as WD\,2151-015B. Again, the size of the offset suggests and the lack of any
radial velocity variation suggests the presence of a distant, cool companion
star.  The radial velocity measurements for WD\,2151-015 are included in our
analysis, but those of its companion are not.} \begin{figure}
\leavevmode\centering{
\psfig{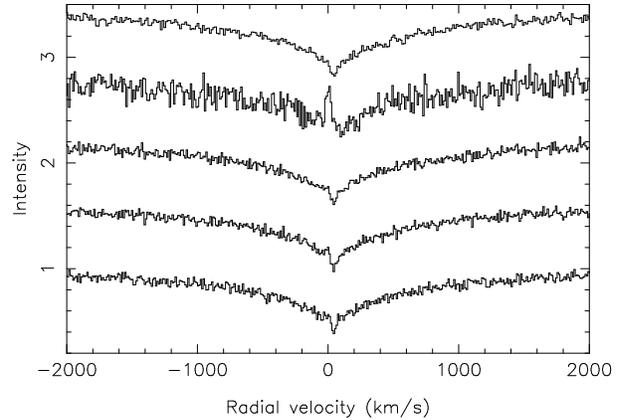} }
\caption{\label{WD2151Fig} Spectra of WD\,2151-015 showing the absorption from
the white dwarf and the variable emission line from the proposed companion
star.} \end{figure} 
\item[\bf WD\,2211-495]{The sharp core of this star is reversed (in emission).
Unlike those stars with companion stars, there is no significant offset
between emission and absorption in this star ($\loa 5\kms$). The emission is
simply a consequence  of the higher temperature of this star ($\sim$65\,000K;
Finley et~al, 1997) compared to others in our sample
(typically $\sim$20\,000K). There appears to be no reason why these emission
cores should not be used to extend radial velocity surveys of DA white dwarfs
into these hotter regimes.}
\item[\bf WD\,2326+049]{This high-amplitude ZZ~Ceti variable star shows a
large infrared excess which continues to defy explanation. The presence of a
companion star is almost certainly ruled out by the lack of any light-time
effect in the pulsation periods (Kleinman et~al., 1994). The favoured
hypothesis now appears to be circumstellar dust, though this requires very
special geometries to explain the variability of the infrared excess on
periods longer than those seen in the star itself (Patterson et~al., 1991). We
have used this star in our analysis, but for consistency have ignored the many
other published radial velocity measurements. }

\item[\bf WD\,2359-434]{There is no obvious sharp core to the \halpha\
line apparent in our spectra of this star (Fig.~\ref{WD2359Fig}).  A single
higher resolution spectrum by Koester shows a very weak, sharp core
unlike those seen in other white dwarfs. There is also some hint of
variability in the \halpha\ line from our spectra. The broad wings
of the \halpha\ line are similar to white dwarfs of similar spectral type.
This odd star clearly deserves further investigation. We used only three
Gaussian profiles to model the profile of this star and have included it in
our analysis.} 
\begin{figure} \leavevmode\centering{
\psfig{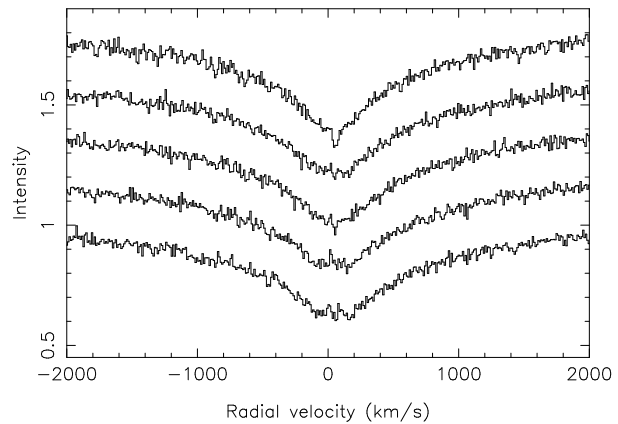} }
\caption{\label{WD2359Fig} Spectra of WD\,2359-434 showing the anomalous core
of \halpha.} \end{figure} 
\end{description}

\section{Results}

\begin{table}
\caption{\label{DDTable} Parameters of known DDs (data from Moran, 1999). The
mass ratio, q, is assumed to be 1 when it is not known.}
\begin{tabular}{lrrrrr}
WD&\multicolumn{1}{c}{P(d)}&\multicolumn{1}{c}{q}  &
\multicolumn{1}{c}{M/\Msolar} & \multicolumn{1}{c}{K(\kms)} &
\multicolumn{1}{c}{P(Detect)} \\
0135$-$052 & 1.556  &  0.90 & 0.47& 78 & 100\% \\
0136$+$768 & 1.407  &  1.31 & 0.34& 65 & 100\% \\
0957$-$666 & 0.061  &  1.14 & 0.37& 219 & 100\% \\
1022$+$050 & 1.157  &       & 0.35& 74 &  93\% \\
1101$+$364 & 0.145  &  0.87 & 0.31& 70  &  100\% \\
1202$+$608 & 1.493  &       & 0.40& 77 &  100\% \\
1204$+$450 & 1.603  &  1.00 & 0.51& 96 &  100\% \\
1241$-$010 & 3.347  &       & 0.31& 68 &  100\% \\
1317$+$453 & 4.872  &       & 0.33& 64   &  93\% \\
1713$+$332 & 1.123  &       & 0.38& 55 & 89\% \\
1824$+$040 & 6.266  &       & 0.39& 59 & 87\% \\
2032$+$188 & 5.084  &       & 0.36& 64 & 93\% \\
2331$+$290 & 0.167  &       & 0.39& 164 &  100\% \\
\end{tabular}
\end{table}

\subsection{The criterion for detection of a binary.}
 The uncertainties on every data point in the spectra are propagated through
the data reduction and analysis and so the measured radial velocities have
accurate uncertainties associated with them. We can, therefore, calculate a
weighted mean radial velocity for the spectra of each star. This mean is the
best estimate of the radial velocity of the star assuming this quantity is
constant. We then calculate the $\chi^2$ statistic for this ``model'', i.e.
the goodness-of-fit of a constant to the observed radial velocities. We can
then compare the observed value of $\chi^2$ with the distribution of
$\chi^2$  for the appropriate number of degrees of freedom. In the case of
significant variability in the radial velocities the probability of obtaining
the observed value of $\chi^2$ or higher from random fluctuations of constant
radial velocities, $p$,  will be extremely low. The observed values of the
weighted mean radial velocity, $\chi^2$ and the logarithm of this probability,
$\log_{10}p$, are given for all the binaries in our sample in
Table~\ref{ResultsTable}.

 To test whether the uncertainties for our measured radial velocities are
reliable we calculated the total value of $\chi^2$ for all the stars in our
sample, $\chi_{\rm tot}^2 = 180.3$,. The number of degrees of freedom, $\nu$,
is given by $\nu = n_{\rm rv} - n_{\rm obj} = 214 - 48 = 166$, where $n_{\rm
rv}$ is the number of radial velocity measurements and $n_{\rm obj}$ is the
number of stars measured (including WD\,1845+019\,B and WD\,2151-015\,B). For
a $\chi^2$ distribution with $\nu=166$ we find the probability ${\rm P}(\chi^2
> 180.3) = 0.21$. If almost all the stars in the sample are non-variable this
suggests that our uncertainties are very reliable. 

  We have decided to use $\log_{10}p < -3$ as the criterion for detection of a
binary, i.e. a 1/1000 probability of random fluctuations given the observed
value of $\chi^2$ or higher.  This is a sufficiently low probability that the
chances of one or more spurious detections in our sample is less than 5\%, but
sufficiently high for genuine binaries not to be missed. The latter point is
demonstrated in Table~\ref{DDTable} where we give the probability of 
detecting the binary nature of some known DDs, P(Detect), given  a typical
observing sequence of 4 spectra yielding radial velocities with uncertainties
of 3\kms. We see that the majority of known DDs would never be missed and that
even those with long or awkward periods have a good chance of being detected. 

Reference to Table~\ref{ResultsTable} shows that there no stars in our sample
which are binaries according to our detection criterion. However, there are
two stars that were included in our observing list which we deliberately did
not observe as they were known to be binaries prior to our observations,
namely  WD\,0135+052 and WD\,1824+040. By examination of the magnitude and RA
distributions of the stars we did observe it is clear that these stars would
certainly have been observed had they not been known to be binaries and they
would almost certainly have been detected as binary using our observing
strategy (Table~\ref{DDTable}).  This is a rather unsatisfactory situation
which we can only resolve easily by including these two binaries in our
analysis as though they had been observed as part of the main sample and
assuming they would have been detected as binaries.

\subsection{The fraction of double degenerates among DA white dwarfs.}
 Given a theoretical model which predicts the period, mass and mass ratio
distributions of DDs, we can calculate the probability of obtaining the
observed number of binaries and their period distribution i.e. we can
calculate the probability of obtaining the data, D, given the model, M, and
the binary fraction for DDs, $f$, P(D\,$|$\,M,$f$). This is demonstrated in
Appendix A. To find the probability distribution of $f$, we appeal to Bayes'
theorem to show that
\[ \rm{P(f\,}|{\rm \,M,D)} = \frac{{\rm P(D\,}|{\rm \,M,f)P(f\,}|{\rm \,M)}}
{{\rm P(D\,}|{\rm \,M)}} \] 
By assuming the model to be correct, i.e. $\int_0^1$ P(f\,$|$\,M,D)$df=1$, and
assuming that all values of $f$ are equally likely (i.e. the prior probability
P(f\,$|$\,M) is constant), we can simply normalise the P(D\,$|$\,M,f) to get
P(f\,$|$\,M,D), the probability distribution of $f$ given the data and the
model. Simple numerical integration then enables us to find a range of values
which gives some probability of including the true value of $f$.

 We calculated the probable range of $f$ for two theoretical models,
Iben et~al., 1997 and Han, 1998. The first step is to calculate the average
detection efficiency as a function of period, $d(p)$ i.e. the probability of
detecting a binary at a given period using the measured radial velocities with
the adopted detection criterion and including the effects of randomly oriented
orbits. For the model of Iben et~al. we calculated  the mass of the brighter
white dwarf, $M_1$, using $\log(M_1) = 0.13 \log(P) -0.6$, where the mass is in
solar masses and the period, $P$, is in days. This is simply an approximation to the
main feature of the bivariate distribution of periods and masses for DDs given
by Saffer et~al. (1998) for the same model. The model of Han shows the mass to
be almost independent of period and to have a bi-modal distribution.
Therefore, we took the average of $d(p)$ for $M_1=0.3$ with weight of 1 and
$M_1=0.6$ with a weight of 0.7.  The mass ratio distribution for the two
models is similar and is strongly peaked so we used a single, average value
for the mass ratio of 0.67 for both. We used the data for the 46 white dwarfs
presented in Table~\ref{RVTable} to calculate $d(p)$ for each model at 2500
periods and then summed into histogram with 50 period bins before forming the
average. 

\begin{figure}
\leavevmode\centering{\psfig{file=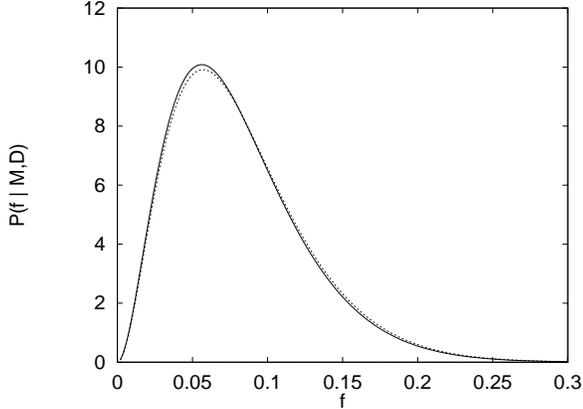,width=0.45\textwidth} }
\caption{\label{FdistFig} The probability distribution of $f$ assuming a the
model of Han (dotted line) or Iben et~al. (solid line) to be true and with a
uniform prior probability distribution assumed for $f$. }
\end{figure}

\begin{figure}
\leavevmode\centering{\psfig{file=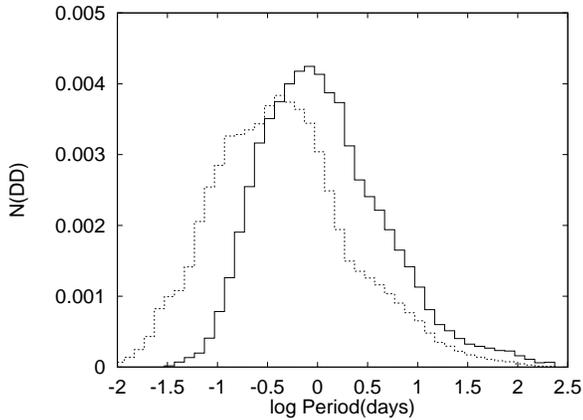,width=0.45\textwidth} }
\caption{\label{NBFig} The predicted period distribution of DDs in our
sample from the model of Iben et~al. (solid line) and Han (dotted line)
integrated over the probability distribution of $f$.} 
\end{figure}

  The probability distributions P(f\,$|$\,M,D) for the models of Han and Iben
et~al. are shown in Fig.~\ref{FdistFig}. We find that the fraction of DDs
among DA white dwarfs has a 95\% probability of lying in the range
$[0.017,0.19]$ independent of the model used. The predicted period
distribution of binaries in our average detection efficiency and sample size
for each model averaged over the distribution P(f\,$|$\,M,D)  is given in
Fig.~\ref{NBFig}.

\subsection{Agreement between observations and theory.}
 To get an idea of how well the observations agree with the two proposed
theories, we compared the probability of obtaining the data given each model
integrated over $f$, $\int_0^1$\,P(D\,$|$\,M,f)\,$d$f, to the same quantity
calculated for a simple default model. For this default model we assumed that
the periods are uniformly distributed between 0.06\,d and 300\,d. The lower
limit is set by the period below which two white dwarfs with the typical mass
of those in our survey (0.5\Msolar) would merge within their typical cooling
age ($10^8$\,y). The upper age is set by the maximum period for DD resulting
from a common envelope phase estimated from Fig~5 of Saffer et~al. The exact
value of these limits has little effect on the results. The mass and mass
ratio distributions for the simple model were taken to be the same as those
for the model being tested in each case. We find that the model of Han is 3.3
times more likely to produce the observed number and period distribution of
binaries than the simple model. The equivalent factor for the model of Iben
et~al. is 1.4. Thus, both models are no worse than a simple model of the
period distribution.

\section{Discussion}
 We have been able to perform a quantitative analysis of a survey for binary
stars among DA white dwarfs. Despite the relatively large sample of stars and
the high detection effeciencies for binaries in the expected period range, we
are only able to measure the fraction of DDs among DA white dwarfs to within a
factor of ten. To reduce this uncertainty to a factor of three would require a
sample approximately 4 times larger than the one presented here, with a
similar detection efficiency (or a much larger survey with a lower detection
efficiency). 

\begin{figure}
\leavevmode\centering{
\psfig{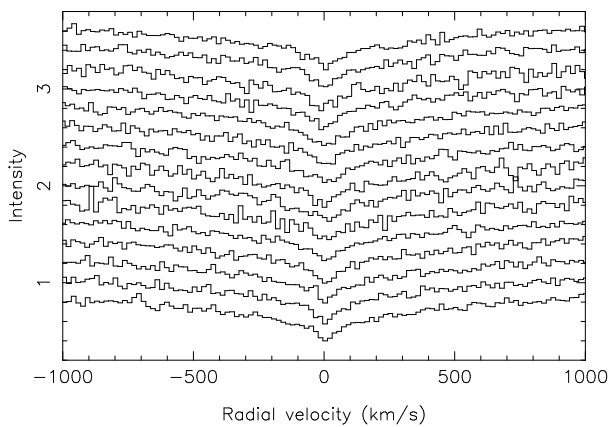} } 
\caption{\label{WD1310Fig} Spectra of WD\,1310+583 obtained during 1997 June 
22\,--\,23 and 1998 February 8\,--\,10.} 
\end{figure}

 The two  binaries in our sample of 48 stars contrasts sharply with 
the survey of Saffer et~al. (1998), who claim to have found 13 new binaries in a
survey of 107 stars. Although consistent with our range of $f$, this is a
little puzzling given that their survey observed each star only three times
and with a lower precision per measurement, both of which imply a lower
detection efficiency than ours for the majority of DDs.  However, this does
not take account of the ``false alarm'' rate for Saffer et~al.'s survey.
Including false detections in the analysis is rather problematic, so our
detection criterion was deliberately choosen so as to keep this rate low
enough to give a less than 5\% chance of one or more false detections. By
contrast, if we take the uncertainty of 25\kms\ quoted by Saffer et~al. for
their radial velocity measurements and their quoted detection criterion of any
two spectra differing by 65\kms, we find that their sample should contain an
average of 17.7 false detections, which would suggest that on their figures,
{\it none} of their candidates are likely to be genuine binaries! This is
clearly not reasonable given that 7 of their candidates were known to be
binaries beforehand or have been confirmed by follow-up observations and have
measured orbital periods. On the other hand, observations similar to those
presented here have also shown that at least two of the candidates are
probably single. We have observed several of Saffer et~al.'s candidate
binaries using the 4.2m William Herschel and 2.5m Isaac Newton telescopes  on
the Island of La Palma in the Canary Islands. Our 10 spectra of WD\,1232+479,
one of Saffer et~al.'s ``weight~1'' candidates, obtained over 3 nights have a
typical uncertainty of 4\kms and show no sign of variability ($\log_{10}p =
-0.59$). Similarly, we obtained 15 spectra of WD\,1310+583, another of Saffer
et~al.'s ``weight~1'' candidates, during two observing runs 8 months apart,
over two and three nights and, again, found no variability ($\log_{10}p =
-0.04$). Inspection of the spectra of WD\,1310+583 (Fig.~\ref{WD1310Fig})
shows that there is certainly no variation as large as the $\sim 100\kms$
shift that is seen in the spectra of Saffer et~al. (their Fig.~1). If there
are only 2 false alarms among the ``weight~1'' candidates of Saffer et~al.,
their radial velocities are accurate to $\sim 17\kms$. However, the false
alarm rate is very sensitive to the exact value of the uncertainty. We have
also assumed that the detection criterion given by Saffer et~al. is an
accurate reflection of the subjective methods they actually used. These are
impossible to quantify in practice and so it is extremely difficult to draw
any concusions concerning the statistics of DDs from their survey. 

 We also analysed the radial velocities given by Foss et~al. for 25 DA white
dwarfs. Their radial velocities are typically ten times less accurate than
ours and so their data alone provide a very poor constraint on the  binary
fraction for DDs. If we incorporate their results into our survey we find this
makes essentially no difference to our results. This is simply because radial
velocities surveys of such low accuracy can only detect very short period
binaries, and there are expected to be very few such systems in the sample
of all known white dwarfs and probably none at all in such a small sample.

 Given the large uncertainty in our estimate of the binary fraction, it is
obvious that there is insufficient data at present  to determine whether
merging DDs are a viable source of Galactic SNe~Ia. This will remain the case
until a much larger sample of white dwarfs is surveyed for DDs. There are also
other issues which must be addressed before we answer this question. The
validity of the theoretical models can be tested using techniques similar to
those presented here if a new, large survey is also designed to be quantative
from the outset. Such a survey should also consider the selection effects in
the sample of stars to be observed, particularly with regard to known
binaries, in a more careful manner than has been done here. The space density
of white dwarfs, the luminosity function and the number of very cool halo
white dwarfs are also issues which will need to be tackled before we can come
to a definite conclusion on this point.

\section{Conclusion}
 We have used a quantitative radial velocity survey of 46 stars with 
a high detection efficiency to measure the fraction of double degenerates
among DA white dwarfs. We find a 95\% probability that the  lies in the range
[0.017,0.19], independent of the details of the model used to predict the
period, mass and mass ratio distribution.  The combination of sample size and
high radial velocity accuracy make this the most sensitive survey for double
degenerates with periods of hours and days conducted to date. The rather loose
constraint on the binary fraction we derive illustrates the need to
increase the number of white dwarfs observed. Such a survey will need to be
quantitative if an objective detection criterion for binaries is to be used.
This would appear to be essential if a valid comparison is to be made between
observations and theory.

\section*{Acknowledgements}
TRM was supported by a PPARC Advanced fellowship. 
PFLM was supported on a PPARC post-doctoral grant.
We would like to thank Rex Saffer for making the results of his survey
available to us.

\setcounter{table}{0}
\begin{table*}
\caption{\label{RVTable} Radial velocity measurements for our
sample of DA white dwarfs.}
\begin{tabular}{lrrr|lrrr}
WD &  \multicolumn{1}{l}{HJD} &\multicolumn{1}{l}{Radial velocity} & 
WD &  \multicolumn{1}{l}{HJD} &\multicolumn{1}{l}{Radial velocity} \\
     &  \multicolumn{1}{l}{-2450000} & (\kms)  &
     &  \multicolumn{1}{l}{-2450000} & (\kms)  \\
0047-524 & 677.1278 & 31.6 $\pm$ 2.5 & 1550+183 & 970.0070 & 15.0 $\pm$ 3.8 \\ 
 & 677.1385 & 34.7 $\pm$ 2.5 &  & 970.0178 & 14.8 $\pm$ 3.5 \\  
 & 677.3004 & 39.2 $\pm$ 2.5 &  & 970.1484 & 17.4 $\pm$ 3.5 \\  
 & 968.3358 & 35.2 $\pm$ 3.1 &  1616-591 & 968.1108 & 10.4 $\pm$ 4.1 \\ 
 & 969.3412 & 38.8 $\pm$ 3.0 &  & 968.1250 & 8.8 $\pm$ 3.8 \\ 
0101+048 & 677.1806 & 63.6 $\pm$ 2.5 & & 968.9571 & 14.0 $\pm$ 5.1 \\  
 & 677.1879 & 62.7 $\pm$ 2.7 &  & 969.1295 & 6.2 $\pm$ 4.1 \\ 
 & 677.3149 & 65.4 $\pm$ 3.6 &  1619+123 & 968.0131 & 20.2 $\pm$ 4.2 \\ 
0226-329 & 676.2976 & 30.4 $\pm$ 2.4 & & 968.0273 & 29.4 $\pm$ 4.6 \\  
 & 676.3049 & 23.3 $\pm$ 2.4 &  & 968.1627 & 23.5 $\pm$ 5.1 \\  
0227+050 & 676.2800 & 19.9 $\pm$ 1.0 & & 968.9742 & 19.8 $\pm$ 3.2 \\  
 & 676.2873 & 18.2 $\pm$ 1.1 &  1620-391 & 676.8555 & 45.2 $\pm$ 0.8 \\ 
 & 677.1941 & 18.2 $\pm$ 2.4 &  & 676.8593 & 46.4 $\pm$ 0.9 \\  
 & 677.3234 & 17.3 $\pm$ 2.1 &  & 676.8631 & 47.0 $\pm$ 0.9 \\  
0310-688 & 676.3119 & 65.9 $\pm$ 0.7 & & 677.0417 & 45.0 $\pm$ 1.9 \\  
 & 676.3157 & 66.2 $\pm$ 0.7 &  & 677.0435 & 44.8 $\pm$ 2.1 \\  
 & 676.3195 & 65.5 $\pm$ 0.8 &  & 677.8541 & 47.6 $\pm$ 1.0 \\  
 & 676.3234 & 66.5 $\pm$ 0.7 &  & 968.0005 & 47.5 $\pm$ 1.5 \\  
 & 676.3272 & 65.3 $\pm$ 0.7 &  1659-531 & 967.9799 & 50.8 $\pm$ 1.8 \\ 
 & 676.3317 & 65.0 $\pm$ 0.8 &  & 967.9907 & 51.7 $\pm$ 2.2 \\  
 & 676.3355 & 65.2 $\pm$ 0.8 &  & 968.1790 & 51.4 $\pm$ 1.9 \\  
 & 676.3396 & 66.7 $\pm$ 0.9 &  1716+020 & 676.9091 & -16.2 $\pm$ 2.2 \\ 
 & 677.2001 & 66.4 $\pm$ 1.5 &  & 677.0235 & -18.9 $\pm$ 3.0 \\  
 & 677.2025 & 67.1 $\pm$ 1.6 &  & 677.0343 & -15.7 $\pm$ 3.0 \\  
 & 677.3308 & 69.2 $\pm$ 1.9 &  & 677.8681 & -15.3 $\pm$ 1.6 \\  
 & 677.3328 & 62.4 $\pm$ 2.1 &  & 968.1461 & -12.9 $\pm$ 3.0 \\  
 & 677.3349 & 67.1 $\pm$ 2.1 &  1743-132 & 969.0931 & -67.6 $\pm$ 5.3 \\ 
 & 677.3370 & 64.9 $\pm$ 2.8 &  & 969.1003 & -66.8 $\pm$ 4.9 \\  
1149+057 & 969.8734 & 2.0 $\pm$ 6.0 & & 970.0914 & -68.3 $\pm$ 3.5 \\  
 & 969.8842 & 0.6 $\pm$ 6.4 & & 970.2120 & -67.3 $\pm$ 3.4 \\  
 & 970.0278 & 4.0 $\pm$ 8.0 & & 970.2193 & -70.8 $\pm$ 3.4 \\  
1210+140 & 969.8965 & 65.2 $\pm$ 5.4 & 1826-045 & 968.0441 & 1.1 $\pm$ 2.6 \\ 
 & 969.9073 & 74.8 $\pm$ 5.8 &  & 968.0583 & 3.2 $\pm$ 2.8 \\ 
1233-164 & 969.9216 & 75.1 $\pm$ 10.2 & & 969.0310 & -0.5 $\pm$ 2.1 \\ 
 & 969.9324 & 64.5 $\pm$ 7.6 &  & 969.2441 & 3.1 $\pm$ 1.5 \\ 
 & 970.0424 & 63.0 $\pm$ 7.5 &  1827-106 & 969.1088 & -33.0 $\pm$ 7.0 \\ 
1310-305 & 968.8958 & 41.6 $\pm$ 3.4 & & 969.1161 & -35.5 $\pm$ 7.0 \\  
 & 969.9612 & 42.2 $\pm$ 2.7 &  & 970.1005 & -29.1 $\pm$ 3.9 \\  
 & 970.0674 & 37.7 $\pm$ 2.1 &  1840+042 & 970.1647 & 4.3 $\pm$ 2.2 \\ 
 & 970.0782 & 40.4 $\pm$ 2.1 &  & 970.1754 & 6.2 $\pm$ 2.2 \\ 
1314-153 & 968.9897 & 101.0 $\pm$ 4.1 & & 970.2672 & 6.7 $\pm$ 2.2 \\ 
 & 969.0128 & 109.9 $\pm$ 3.9 &  1840-111 & 968.0754 & -6.5 $\pm$ 3.3 \\ 
 & 969.9472 & 113.6 $\pm$ 2.9 &  & 969.0474 & -3.8 $\pm$ 2.5 \\  
 & 970.0553 & 105.4 $\pm$ 2.3 &  & 969.0616 & -7.1 $\pm$ 2.2 \\  
 & 970.1213 & 111.3 $\pm$ 2.5 &  & 969.2600 & -5.8 $\pm$ 2.2 \\  
1327-083 & 967.9596 & 43.9 $\pm$ 1.2 & 1845+019 & 676.8921 & -32.6 $\pm$ 2.0 \\ 
 & 967.9653 & 43.6 $\pm$ 0.9 &  & 676.8994 & -27.9 $\pm$ 1.9 \\  
 & 968.0883 & 44.2 $\pm$ 1.1 &  & 677.0960 & -31.4 $\pm$ 3.2 \\  
 & 968.8552 & 45.3 $\pm$ 0.7 &  & 677.1010 & -21.2 $\pm$ 4.4 \\  
 & 969.8351 & 45.2 $\pm$ 0.8 &  & 677.9255 & -30.6 $\pm$ 2.0 \\  
1348-273 & 969.9835 & 62.4 $\pm$ 5.5 & 1845+019B & 676.8921 & -53.3 $\pm$ 2.2 \\
 & 969.9943 & 59.3 $\pm$ 6.2 &  & 676.8994 & -48.4 $\pm$ 2.1 \\  
 & 970.1359 & 63.8 $\pm$ 4.1 &  & 677.0960 & -53.6 $\pm$ 3.8 \\  
1425-811 & 675.8778 & 32.4 $\pm$ 2.2 & & 677.1010 & -39.8 $\pm$ 4.7 \\  
 & 675.8850 & 33.0 $\pm$ 2.2 &  & 677.9255 & -47.1 $\pm$ 2.0 \\  
 & 676.8715 & 37.2 $\pm$ 3.0 &  1914-598 & 675.8993 & 74.5 $\pm$ 2.1 \\ 
 & 676.8788 & 33.3 $\pm$ 3.2 &  & 676.0923 & 74.6 $\pm$ 2.4 \\  
 & 677.0067 & 37.5 $\pm$ 2.3 &  & 676.1030 & 75.3 $\pm$ 2.2 \\  
 &  &  &  & 677.9536 & 75.3 $\pm$ 3.7 \\  
\end{tabular}  
\end{table*}
\setcounter{table}{0}
\begin{table*}
\caption{continued.}
\begin{tabular}{lrrr|lrrr}
WD &  \multicolumn{1}{l}{HJD} &\multicolumn{1}{l}{Radial velocity} &
WD &  \multicolumn{1}{l}{HJD} &\multicolumn{1}{l}{Radial velocity} \\
     &  \multicolumn{1}{l}{-2450000} & (\kms)  &
     &  \multicolumn{1}{l}{-2450000} & (\kms)  \\
1919+145 & 676.9222 & 52.3 $\pm$ 1.5 & 2159-754 & 970.2918 & 153.1  $\pm$ 3.1 \\
 & 676.9294 & 55.3 $\pm$ 1.4 &  & 970.3026 & 153.2  $\pm$ 3.4 \\
 & 677.0896 & 53.6 $\pm$ 2.1 & 2211-495 & 968.3460 & 38.2  $\pm$ 5.3 \\ 
 & 677.9670 & 51.2 $\pm$ 3.7 &  & 968.3515 & 32.0  $\pm$ 7.6 \\ 
 & 677.9785 & 49.1 $\pm$ 2.1 &  & 970.2548 & 35.4  $\pm$ 4.0 \\ 
1943+163 & 969.1790 & 34.3 $\pm$ 4.5 &  & 970.3491 & 41.7  $\pm$ 4.1 \\ 
 & 969.1863 & 34.8 $\pm$ 4.3 &  & 970.3529 & 36.4  $\pm$ 4.2 \\ 
 & 970.1076 & 36.3 $\pm$ 2.4 & 2251-634 & 677.0781 & 29.5  $\pm$ 2.4 \\ 
 & 970.2355 & 37.6 $\pm$ 2.1 &  & 677.2156 & 35.5  $\pm$ 3.0 \\ 
2007-303 & 675.9451 & 75.1 $\pm$ 0.7 &  & 677.2298 & 30.9  $\pm$ 3.0 \\ 
 & 676.1501 & 73.5 $\pm$ 0.7 &  & 968.2887 & 33.4  $\pm$ 2.7 \\ 
 & 676.1574 & 75.7 $\pm$ 0.7 &  & 969.3024 & 33.9  $\pm$ 2.5 \\ 
 & 968.1908 & 78.4 $\pm$ 1.1 & 2326+049 & 676.0558 & 44.7  $\pm$ 1.7 \\ 
 & 969.2190 & 77.0 $\pm$ 1.0 &  & 676.0631 & 41.5  $\pm$ 1.8 \\ 
2014-575 & 675.9178 & 43.2 $\pm$ 5.0 &  & 676.2597 & 43.8  $\pm$ 1.8 \\ 
 & 675.9347 & 39.1 $\pm$ 4.0 &  & 968.3044 & 48.3  $\pm$ 1.8 \\ 
 & 676.1397 & 40.5 $\pm$ 3.2 &  & 969.3476 & 49.3  $\pm$ 5.0 \\ 
 & 678.0281 & 43.2 $\pm$ 5.1 & 2333-165 & 676.0334 & 72.3  $\pm$ 1.2 \\ 
2035-336 & 676.9718 & 21.3 $\pm$ 3.5 &  & 676.2437 & 72.6  $\pm$ 1.1 \\ 
 & 676.9825 & 22.3 $\pm$ 2.9 &  & 676.2510 & 72.2  $\pm$ 1.1 \\ 
 & 677.1613 & 22.5 $\pm$ 3.3 &  & 677.2745 & 71.7  $\pm$ 1.6 \\ 
 & 968.2061 & 18.0 $\pm$ 2.5 & 2351-335 & 676.1180 & 47.9  $\pm$ 1.5 \\ 
 & 969.2275 & 21.7 $\pm$ 2.6 &  & 676.1287 & 53.2  $\pm$ 1.6 \\ 
2039-202 & 676.9419 & -0.0 $\pm$ 2.0 &  & 676.2713 & 49.8  $\pm$ 1.6 \\ 
 & 676.9457 & 0.8 $\pm$ 2.1 &   & 968.3206 & 52.8  $\pm$ 2.1 \\ 
 & 677.1705 & 3.2 $\pm$ 1.8 &   & 969.3128 & 52.8  $\pm$ 2.1 \\ 
 & 677.1743 & 0.6 $\pm$ 1.8 &  2359-434 & 676.0161 & 51.9  $\pm$ 3.6 \\ 
 & 678.1085 & 0.2 $\pm$ 1.7 &   & 676.0234 & 48.7  $\pm$ 3.6 \\ 
 & 968.2196 & 1.7 $\pm$ 1.0 &   & 676.2279 & 45.7  $\pm$ 3.7 \\ 
 & 969.2692 & 4.6 $\pm$ 1.6 &   & 676.2341 & 37.4  $\pm$ 3.5 \\ 
2039-682 & 677.1090 & 52.6 $\pm$ 4.2 &  & 677.2868 & 37.6  $\pm$ 4.0 \\ 
 & 677.1162 & 60.8 $\pm$ 4.7 &  & 969.3255 & 43.3  $\pm$ 3.8 \\ 
 & 677.2433 & 49.8 $\pm$ 6.1 &  & 969.3328 & 34.7  $\pm$ 5.7 \\ 
2058+181 & 970.1852 & -40.8 $\pm$ 5.7 & & 970.3186 & 43.0  $\pm$ 2.9 \\  
 & 970.1959 & -37.3 $\pm$ 6.3 & \\
 & 970.2770 & -45.5 $\pm$ 3.5 & \\
2105-820 & 676.0043 & 45.2 $\pm$ 2.6 & \\
 & 676.2096 & 43.0 $\pm$ 3.2 & \\
 & 676.2169 & 42.4 $\pm$ 3.2 & \\
 & 968.2334 & 45.9 $\pm$ 3.6 & \\
 & 969.2781 & 37.2 $\pm$ 2.9 & \\
2115-560 & 675.9846 & 17.2 $\pm$ 3.2 & \\
 & 675.9954 & 7.1 $\pm$ 1.7 & \\
 & 676.2005 & 9.6 $\pm$ 2.4 & \\
 & 968.2507 & 9.8 $\pm$ 2.1 & \\
 & 969.2911 & 12.6 $\pm$ 3.5 & \\
2149+021 & 675.9547 & 31.7 $\pm$ 1.1 & \\
 & 675.9620 & 31.1 $\pm$ 1.1 & \\
 & 676.1671 & 32.3 $\pm$ 0.9 & \\
 & 678.1756 & 34.5 $\pm$ 2.5 & \\
2151-015 & 675.9728 & 32.1 $\pm$ 3.2 & \\
 & 676.1780 & 37.9 $\pm$ 2.7 & \\
 & 676.1887 & 30.5 $\pm$ 3.0 & \\
 & 677.2606 & 24.1 $\pm$ 9.1 & \\
 & 968.2692 & 43.0 $\pm$ 3.1 & \\
2151-015B & 675.9728 & 9.1 $\pm$ 13.5 & \\
 & 676.1780 & -5.7 $\pm$ 17.2 & \\
 & 676.1887 & 0.4 $\pm$ 18.0 & \\
 & 677.2606 & 10.1 $\pm$ 3.2 & \\
 & 968.2692 & 18.1 $\pm$ 9.6 & \\
2151-307 & 970.2461 & 53.7 $\pm$ 5.7 & \\
 & 970.3335 & 53.3 $\pm$ 5.8 & \\
 & 970.3425 & 63.3 $\pm$ 9.1 & \\
\end{tabular}
\end{table*}

\newpage

\appendix
\section{The probability distribution of the binary fraction given the
observed data.} 
 In order to calculate the probability distribution of the binary fraction,
$f$, given a survey for binaries, we consider a histogram of observed and
predicted periods with $M$ period bins covering the entire range of possible
periods.  The theoretical model provides the probability of a given binary
having a period within bin $i$, $p_i$, $i=1,\dots ,M$. Since this probability
applies only to binaries and the histogram of predicted periods covers the
entire range of possible periods, $\sum_{i=1}^M p_i = 1$. 

 Now consider a sample of N stars. For a given detection criterion we can
calculate for each star the probability of detecting it to be binary with a
given period. The average probability of detection over all the stars in the
sample integrated over all periods in bin $i$ is $d_i$. For each star we have
$(M+1)$ possible outcomes, M possible periods or a non-detection. Each of the
M possible periods have probabilities $d_i p_i f$. The non-detections have
probabilities $(1-\sum_{i=1}^M d_i p_i f) = (1-\overline{d}f)$, where
$\overline{d} = \sum_{i=1}^M d_i p_i$ is the mean detection probability for
the model distribution. We assume that the periods of all the systems detected
to be binaries are known and that there are no false detections. The number of
stars found to be binaries with periods in a given bin $i$ is $o_i$.  The
probability of obtaining the observed distribution of periods and
non-detections giving the theoretical model, $P_M$, is then given by the
multinomial distribution as follows: 

\[ P_M = \frac{ N! \prod_{i=1}^M(p_i d_i)^{o_i} } 
              {(N - N_B)!\prod_{i=1}^M o_i!  
              } 
               f^{N_B} 
              (1 -  \overline{d}f)^{(N-N_B)}, \]
where $N_B = \sum_{i=1}^M o_i$ is the number of binaries detected. In the case
of no binary detections, this reduces to the simple case
\[ P_M =  (1 -  \overline{d}f)^N \] 
 Given the large number of products and factorials involved it is easier in
practice to calculate $\log P_M$:
\[\begin{array}{lcl}
\log P_M  &=& \log(N!) + \sum_{i=1}^Mo_i\log(p_i d_i)  \\
         && - \log(N-N_B)! -\sum _{i=1}^M\log(o_i !) \\
         &&  + N_B \log(f) + (N-N_B)\log(1 - \overline{d}f)  \\
\end{array}\]

\label{lastpage} 

\begin{thebibliography}{99}
\bibitem{1} Bergeron P., Saffer R.A., Liebert J., 1992, ApJ,  394, 228
\bibitem{2} Bergeron P., Liebert J. Fulbright M.S., 1995, ApJ, 444, 810
\bibitem{3} Bragaglia A., Greggio L., Renzini A., D'Odorico S., 1990, ApJ,
 365, L13
\bibitem{4} Bragaglia A., Renzini A., Bergeron P., 1995, ApJ, 443, 735
\bibitem{5} Finley D.S., Koester D., Basri G., 1997, ApJ, 488, 375
\bibitem{6} Foss D., Wade R.A., Green R.F., ApJ, 374, 281, 1991 
\bibitem{7} Han Z., 1998, MNRAS, 296, 1019
\bibitem{8} Horne K., 1986, PASP, 98, 609
\bibitem{9} Iben Jr. I., Tutukov A.V., Yungelson L.R., 1997, ApJ, 475, 291
\bibitem{10} Kleinman S.J., Nather R.E., Winget D.E., et~al., 1994, ApJ, 436, 
 875
\bibitem{11} Koester D., Dreizler S., Weidemann V., Allard N.F., 1998,
 A\&A, 338, 612
\bibitem{12} Mayor M., Queloz D., 1995, Nature, 378, 355 
\bibitem{13} McCook G.P., Sion E.M., 1998, ApJS, In press. 
\bibitem{15} Moran C. 1999, PhD thesis, University of Southampton, in prep.
\bibitem{18} Nelemans G., Tauris T. M., 1998, A\&A, 335, 85L
\bibitem{19} Robinson E.L., Shafter A.W., 1987, ApJ, 322, 296
\bibitem{21} Patterson J., Zuckerman B., Becklin E.E., Tholen D.J.,
Hawarden T., 1991, ApJ, 374, 330
\bibitem{22} Saffer R.A. Livio M., Yungelson L.R., 1998, ApJ, 502, 394
\bibitem{23} Sarna M.J., Marks P.B., Smith R.C., 1996, MNRAS, 279, 88
\end{thebibliography}
\end{document}